\documentclass[12pt]{iopart}
\usepackage{graphicx}%
\begin{document}
\title{Quasispecies distribution of Eigen model}
\author{Jia Chen and Sheng Li$^*$}
\address{Department of Physics, Shanghai Jiao Tong University,
Shanghai 200240, China} \ead{$^*$lisheng@sjtu.edu.cn}

\begin{abstract}
We study sharp peak landscapes (SPL) of Eigen model from a new
perspective about how the quasispecies distribute in the sequence
space. To analyze the distribution more carefully, we bring forth
two tools. One tool is the variance of Hamming distance of the
sequences at a given generation. It not only offers us a different
avenue for accurately locating the error threshold and illustrates
how the configuration of the distribution varies with copying
fidelity $q$ in the sequence space, but also divides the copying
fidelity into three distinct regimes. The other tool is the
similarity network of a certain Hamming distance $d_{0}$, by which
we can get a visual and in-depth result about how the sequences
distribute. We find that there are several local optima around the
center (global optimum) in the distribution of the sequences
reproduced near the threshold. Furthermore, it is interesting that
the distribution of clustering coefficient $C(k)$ follows
lognormal distribution and the curve of clustering coefficient $C$
of the network versus $d_{0}$ appears as linear behavior near the
threshold.

$\\$

\noindent{\it Keywords\/ Eigen model, Sharp peak landscapes,
Similarity network, Clustering coefficient, Degree distribution }:
\end{abstract}
\maketitle

\section{Introduction}

Eigen model, which first considered that the evolution can be guided other
than by chance, is the first mathematical model that makes Darwin's idea of
mutation and selection able to work in a simple and seemingly "lifeless"
system of chemical reactants \cite{paper29,paper2}. Two main points of the
Eigen model are about the quasispecies which is defined as a stationary
distribution of genetically close sequences, centered around one or several
master sequences, and the existence of an error threshold above which all
information is lost because of accumulating erroneous mutations. The formation
of quasispecies implies that the target of selection in quasispecies theory is
not a single master sequence but a localized distribution in the sequence
space. After Leuth\"{a}usser mapped Eigen model on a well-known system in
statistical mechanics--- the two dimensional Ising system \cite{paper1}, the
study of evolution has become easier and clearer
\cite{paper16,paper20,paper22,paper23,paper24,paper25,paper26}.

Traditionally, people use the concentration of sequence
\cite{paper2,paper11,paper18} or the mean fitness of population
\cite{paper12,paper13,paper14} to measure the error threshold of
evolution. But both of them just divide the copying fidelity into
two regimes for sharp peak landscapes (SPL). They can't
distinguish the area near the threshold from the area far above
the threshold. Moreover, no in-depth results about how the
sequence distribute in the sequence space has been gained in
evolutionary study. Here, we advance a different method by
utilizing the variance of Hamming distance of the sequences at a
given generation to locate the error threshold of SPL. The
simulation shows clearly three distinct regimes of the copying
fidelity.

Today, pairwise sequence comparison is the most widely used application of
bioinformatics because high sequence similarity usually implies structural,
functional or evolutionary relationship among bio-molecular sequences
\cite{paper28}. And the tool of complex network has been already brought in
biological study \cite{paper8,paper9}. It offers unforeseen possibilities to
uncover the organizing principles that govern the formation and evolution of
complex systems. Therefore, we also bring in another tool--- similarity
network to study the Eigen model.

During the evolution, we show that there are two jump
discontinuities in the curve of the average of clustering
coefficient $\left\langle C\right\rangle $ of the network versus
copying fidelity $q$. This phenomenon implies that the phase
transition of the similarity network topology occurs just as we
expected. So we prove that it is logical to study evolution by
bringing in similarity network. We analyze the topology of the
similarity networks established in three different areas of $q$ by
two main basic measures, that is, the degree distribution and the
clustering coefficient. And then we translate the results into
biological language. Thereby, we get a detailed result about how
the sequences distribute in the sequence space. For comparison, we
also present the results of the random sequences created according
to the random selection principle (select 3000 sequences randomly
from $2^{32}$ sequences in the sequence space) in this paper.

This paper is organized as follows. In section $2$, we briefly introduce the
Eigen model. In Section $3$, we introduce the similarity network and explain
how its inherent properties describe the distribution of sequences. In section
$4$, we study SPL to put our ideas in practice and demonstrate how the
sequences generated in SPL distribute in the sequence space. At last we close
our paper with concluding remarks in Section $5$.

\section{The Eigen model and sharp peak landscapes}

Eigen model was established by Eigen in 1971. The original model was aimed at
describing self-replicating RNA or DNA molecules with the assumptions as
follows \cite{paper2}:

\begin{enumerate}
\item The population of evolution is infinite. The individual is represented
by a sequence of $n$ nucleotides $S_{k}=\left\{  s_{1},s_{2},s_{3}%
,...,s_{n}\right\}  $, where $s_{i}$ contains four possible values
to represent the four different bases in the RNA or the DNA
molecules. But for simplicity, people usually take them to be
binary variables which can be considered as distinguishing only
between purines and pyrimidines. Conventionally, $s_{i}$ is
written in the binary form and $S_{k}$ is thus a binary string
$(01000...)$ or equivalent to an integer $b$ $(0\leq b<2^{n})$.

\item Only point mutations are considered. Then the mutation is an exchange of
0 and 1 on a certain locus. For the sake of mathematical tractability, the
point mutation rate $1-q$ ($q$ is the copying fidelity of each position of the
sequence) is constant in time and independent of the position of sequence
(uniform error model).
\end{enumerate}

Years later, Eigen model has been developed to be applicable in a
much wider area. For instance, the results for finite populations
have been gained in both static fitness landscapes
\cite{paper3,paper4,paper7} and dynamic fitness landscapes
\cite{paper5,paper6}. And the error threshold of finite population
$1-q_{f}$ may be lower than the infinite population's
\cite{paper7,paper6}. Thus, for simulated simplicity, we apply
Eigen model to finite population in this report.

The equation describing the reproduction-mutation process is%
\begin{equation}
\frac{dx_{i}}{dt}=\sum_{j=0}^{2^{n}-1}Q_{ij}A_{j}x_{j}-x_{i}\sum_{j=0}%
^{2^{n}-1}A_{j}x_{j} \label{equ1}%
\end{equation}
$x_{i}$ is the concentration of the $i$th sequence $S_{i}$ with normalization
$\sum_{i=0}^{2^{n}-1}x_{i}=1$. $Q_{ij}$ is the probability of getting the
sequence $S_{i}$ as the offspring of the replication of sequence $S_{j}$ and
it is given by%
\begin{equation}
Q_{ij}=q^{n-d_{ij}}(1-q)^{d_{ij}}\label{equ2}%
\end{equation}
where $d_{ij}$ is the Hamming distance between $S_{i}$ and
$S_{j}$. The set of all sequences together with the Hamming
distance forms the sequence space. $A_{i}$ named fitness is the
reproduction rate of $S_{i}$. It means that the sequence $S_{i}$
will reproduce $A_{i}$ offspring every evolution step. The fitness
landscape is then obtained by assigning a numerical value $A_{i}$
to each point in the sequence space \cite{paper15}. The notion of
``fitness landscape" is one of the most powerful concept in
evolutionary theory because evolutionary process could be
considered as a random walk in it.

A broad set of different fitness landscapes has been studied
recently \cite{paper16,paper18,paper12,paper14,paper6,paper15}.
Most of the theoretical work has focused on a particular fitness
landscapes, that is, the sharp peak landscapes
\cite{paper2,paper16,paper20,paper18,paper30}, in which there is a
master sequence $S_{m}=\left\{ s_{1},s_{2},s_{3},...,s_{n}\right\}
$ with larger fitness $A_{m}$, while all other sequences have
uniform fitness (smaller than the master sequence). In
mathematics, it could be expressed as,

\begin{center}
$A_{m}[S_{m}]=A_{0}$ and $A_{i}[S_{i}]=A_{1}<A_{0}$ for any $S_{i}\neq S_{m}$
\end{center}

Ever since Eigen found phase transition in SPL almost 35 years
ago, Eigen's paper has generated substantial interest among the
physicists. The error threshold, a minimal replication accuracy
necessary to maintain the genetic information in the population,
can be viewed as the critical point of a phase transition
separating two regimes of the quasispecies evolution. Above the
error threshold, the distribution of sequences is centered around
the master sequence. Below it, the outcome of a replication event
can be considered a random sequence. This fact is well established
through a great many of works \cite{paper2,paper18,paper30}. And
the exact formulas for the threshold of
ferromagnet and antiferromagnet in the SPL are,%

\begin{equation}
q_{ferror}=(\frac{1}{A_{0}})^{\frac{1}{n}} \label{equ3}%
\end{equation}

\begin{equation}
q_{antiferror}=1-(\frac{1}{A_{1}A_{0}})^{\frac{1}{2n}}\label{equ4}%
\end{equation}
(see \cite{paper2}). The two equations here indicate that the threshold of
ferrormagnet is close to 1 and the antiferromagnetic threshold is close to 0
for large $n$. In order to prepare for studying the SPL of Eigen model, the
similarity network will be introduced in the next section.

\section{Similarity network}

The study of complex networks has been initiated by a desire to understand
various real systems \cite{paper27}. The complex networks can describe a wide
range of systems in nature and society. The direct-viewing and obvious
practical relevance enlighten us to apply this useful tool to the evolutionary
study. A network is composed of a set of nodes and edges
\cite{paper8,paper9,paper19}. In our similarity network, the node i represents
the sequence $S_{i}$ and every node is different from each other. If $d_{ij}$
(the Hamming distance between $S_{i}$ and $S_{j}$) is no larger than the given
Hamming distance $d_{0}$ ($d_{0}\neq0$), we will connect node $i$ to node $j$.
By comparing all the sequences of population, an undirected complex network is
then established \cite{paper10}. It is apparent that the similarity network
will become a complete network if $d_{0}$ is close to $n$ and there will be
little edges in the network if $d_{0}$ is close to $0$. The similarity network
actually reflects the relationships between the sequences of population. It
makes the study of the distribution of sequences easier and more vivid.

Generally speaking, the topology of a network can be quantified by
two main basic quantities. One quantity is the degree
distribution, $P(k)$, giving the probability that a selected node
has exactly $k$ links. The degree distribution allows us to
distinguish among different classes of networks. It can offer us
the approximate configuration about how sequences distribute in
the sequence space. For example, it helps us to know whether the
distribution is homogenous or not in the sequence space. The other
one is the clustering coefficient, a measure of the cliquishness
of a typical neighbourhood, defined as
$C_{i}=2n_{i}/k_{i}(k_{i}-1),$ where $n_{i}$ is the number of
links among the $k_{i}$ nearest neighbors of node $i$. People
usually use the function $C(k)$ and clustering coefficient $C$ of
the network to measure the network's structure. $C(k)$ is defined
as the average clustering coefficient of all nodes with $k$ links.
$C$ is the mean clustering coefficient over all nodes in the
network ($C=\left\langle C_{i}\right\rangle )$, characterizing the
overall tendency of nodes to form clusters. So the second quantity
can picture the compactibility and the local property of the
distribution of sequences. Therefore, we use similarity network to
illustrate in detail how the sequences distribute in the sequence
space and what the relationship is among the sequences of
population by calculating the degree distribution and clustering
coefficient in the coming section.

\section{Quasispecies distribution of sharp peak landscapes}

In this section, we will study the simple model, SPL, to put our
ideas in practice. The error threshold of SPL will be located by
utilizing the variance of Hamming distance of the sequences at a
given generation. Besides, the distribution of sequences
reproduced in this case will be characterized by analyzing the
topology of the similarity network.

On computer simulation aspect, we use a population of $3000$
sequences with $n=32,A_{0}=10,A_{1}=1.$ The total number of
population, the length and the fitness of sequence are invariant
in the process of simulation. The initial condition satisfies that
the homogeneous population is made up of 3000 zeros at $t=0$.
Because the sequence $S_{i}$ will reproduce $A_{i}$ offspring
every evolution step, we need to choose 3000 sequences with the
probability $n_{i}/\sum_{i}n_{i}$ from the offspring after one
replication step in order to keep the total number of sequences
fixed, where $n_{i}$ is the number of sequence $S_{i}$ and
$\sum_{i}n_{i}$ is the total number of the offspring.

Now we begin to carry out our ideas. Firstly, we calculate the
variance of Hamming distance of the sequences at a given
generation to locate the error threshold and characterize how the
configuration of the distribution varies with $q$ in the sequence
space, on the whole.

Figure 1 shows the variance of Hamming distance $\left\langle
var(d_{ij})\right\rangle $ as a function of $q$ for SPL where
$\left\langle ...\right\rangle $ represents the average over 1000
evolution steps from $t=9000$ to $t=10000$ for the given $q$. It
has been proved that it is long enough for the evolution in the
SPL to balance at
t=9000 for the $q$ chosen in figure 1. $var(d_{ij})=\left\langle d_{ij}%
^{2}\right\rangle _{t=t_{0}}-\left\langle d_{ij}\right\rangle
_{t=t_{0}}^{2}$ where $\left\langle ...\right\rangle _{t=t_{0}}$
represents statistical average over different kinds of the
sequences at a given generation $t=t_{0}$. The peaks of the curve
in figure 1 locate at $q=0.9355$ and $q=0.0336$, according well
with the the theoretical values of threshold for the corresponding
parameters, $q_{ferror}=0.9306,q_{antiferror}=0.0353$. The
deflection of the threshold between theory and simulation may be
caused by the finite size effect. The jump discontinuities locate
at $q=0.934$ and $q=0.0344$.

Let us take the ferromagnetic phase ($q>0.5$) as an example to discuss the
sharp peak landscapes. The curve of the ferromagnetic phase in figure 1 is
distinctly made up of three portions, that is, the area above the threshold,
near the threshold and below the threshold. In the area far below the
threshold, the curve of $\left\langle var(d_{ij})\right\rangle $ varying with
$q$ approximates to a horizontal line and its value is equal to $\frac{n}{4}.$
It indicates that the configuration of the distribution far below the
threshold is independent of $q.$\ In the area near the threshold, there is a
peak and its amplitude will become large with the increase in $n$. The
existence of the peak reflects that the scattered band of the distribution is
wide and how complicated the distribution of the sequences generated in this
landscapes is. In the area far above the threshold, $\left\langle
var(d_{ij})\right\rangle $ is close to 0. It says that the distribution of
sequences will converge in the sequence space if $q$ approaches to 1.

Secondly, it is proved that it is logical to study the Eigen model
by bringing in similarity network.

Figure 2 shows $\left\langle C\right\rangle $ as a function of $q$
for SPL where $\left\langle ...\right\rangle $ represents the
average over 1000 evolution steps from $t=9000$ to $t=10000$ of
the similarity network with $d_{0}=8$ for a given $q.$ In figure
2, there are two jump discontinuities at $q=0.934$ and $q=0.0344$,
the same as those in figure 1. We also have analyzed the other
similarity networks with different Hamming distance $d_{0}.$ Most
of them give the same result when $d_{0}$ is not very large. This
phenomenon confirms that the phase transition of the similarity
network topology indeed occurs during the evolution just as we
expected. So it is logical to study evolution by bringing in
similarity network.

Thirdly, the distributions of the sequences generated in three
areas are measured in depth by establishing similarity network.
Furthermore, the results of random sequences are presented for
further analysis.

Figure 3 and figure 4 show respectively the degree distribution and the
function $C(k)$ of similarity networks for $q=0.9355$ (at the threshold) and
$q=0.6$ (far below the threshold). Figure 6(a) and (b) show $C$ as a function
of $d_{0}$ for the similarity networks established by the sequences generated
at the threshold and by the random sequences, respectively.

Let us move on to analyze the topology of the similarity network
to interpret the result of SPL. Near the threshold, $\left\langle
C\right\rangle $ is large (see figure 2). It shows that the
similarity network of this case is compact, namely, the outcomes
of the evolution in this case are genetically close to each other.
In figure 3, there is a relatively small number of nodes which
possess a large number of links (large degree) but not large
clustering coefficient ($C(k)$ is not large). They are known as
hubs, which play the important role of bridging the many small
communities of clusters into a single, integrated network. The
presence of hubs suggests that the master sequence, the center in
the distribution, exists assuredly in the quasispecies. Moreover,
the degree distributions are oscillating in figure 3(a) and (b)
and the oscillation amplitude will become small if the ratio of
$A_{0}$ to $A_{1}$ increases. This phenomenon characterizes that
there are several local optima around the center (global optimum)
in the distribution of the sequences reproduced in the area near
the threshold. It is more interesting that the curve of $C(k)$
follows lognormal distribution (see figure 3(c) and (d)) and the
curve of $C$ versus $d_{0}$ appears as linear behaviour (see
figure 6(a)).

Far below the threshold, $\left\langle C\right\rangle $ is small
(see figure 2) and the curve of $\left\langle C\right\rangle $
varying with $q$ approximates
to a horizontal line just as the curve of $\left\langle var(d_{ij}%
)\right\rangle $ varying with $q$ dose. The degree distribution of the
similarity network with $d_{0}=10$ follows Gauss distribution in figure 4(a)
and the curve of $C(k)$ of the same similarity network is close to a
horizontal line in figure 4(b). For deeper discussion, we introduce the random
sequences created according to the random selection principle (select 3000
sequences randomly from $2^{32}$ sequences in the sequence space). Figure 5
shows the simulations of random sequences. For random sequences, the value of
$\left\langle var(d_{ij})\right\rangle $ could be obtained as follows. We
consider the special case: since the random sequences distribute evenly in the
sequence space, it is assumed that there are $2^{n}$ sorts of sequences in the
population ($n$ is the length of sequence). So,%
\begin{eqnarray}
var(d_{ij}) &  =\left\langle d_{ij}^{2}\right\rangle -\left\langle
d_{ij}\right\rangle ^{2}\label{equ5}\\
&  =\frac{2^{n}(2^{n}-n-1)}{\left(  2^{n}-1\right)  ^{2}}\frac{n}{4}\nonumber
\end{eqnarray}
when n is large enough the equation (\ref{equ5}) approximates to $\frac{n}%
{4}.$ In figure 1, the simulation of $\left\langle var(d_{ij})\right\rangle $
of the sequences created in SPL with $q=0.6$ is equal to $\frac{n}{4}$
(correspond with the random sequences). In figure 5(a), the degree
distribution of the similarity network with $d_{0}=10$ follows Gauss
distribution, the same as that in figure 4(a). In figure 5(b), the curve of
$C(k)$ of the same similarity network approximates to a horizontal line, the
same as that in figure 4(b). All results indicate that both of the two
similarity networks possess the random network topology. So, the outcomes of
the evolution far below the threshold in SPL are equivalent to the random
sequences distributing evenly in the sequence space.

Far above the threshold, the similarity network is very compact which
approximates to a complete network as $\left\langle C\right\rangle $ is close
to 1 no matter what the value of $d_{0}$ is ($d_{0}\neq0$) in this case (see
figure 2), implying that the sequences are highly similar to each other. The
degree distribution and the function $C(k)$ also give the same result.

To sum up, far below the threshold, the values of both
$var(d_{ij})$ and $\left\langle C\right\rangle $ are independent
of $q$. The distribution is uniform in the sequence space. Near
the threshold, the sequences have a large variance of Hamming
distance $var(d_{ij})$ (see figure 1) and a large clustering
coefficient $C$ of the network (see figure 2), which implies that
the scattered band of the distribution is wide and the cliques
have formed in the distribution of sequences, respectively. So, we
conclude that there are several local optima around the center
(global optimum) in the sequence space. Far above the threshold,
the sequences have a small variance of Hamming distance
$var(d_{ij})$ (see figure 1) but a large clustering coefficient
$C$ of the network (see figure 2). So the range of the
distribution will reduce to one point in the sequence space if q
approaches to 1. Finally, it is more interesting that the curve of
$C(k)$ follows lognormal distribution (see figure 3(c) and (d))
and the curve of $C$ versus $d_{0}$ appears as linear behaviour
(see figure 6(a)) near the threshold. By the way, the results of
antiferromagnet are the same as that in the ferromagnetic phase.

\section{Conclusions}

In this paper, we have advanced two ways to study SPL from a
different perspective. One way is the curve of $\left\langle
var(d_{ij})\right\rangle $ versus $q$. It offers us a different
avenue for accurately locating the error threshold. Besides, this
useful means not only shows how the configuration of the
distribution varies with $q$ in the sequence space, but also
divides the copying fidelity into three distinct regimes and makes
a difference between the areas near the threshold and far above
the threshold. The other one is the similarity network, offering a
new avenue to describe the evolution. We have shown that there is
a phase transition in the curve of $\left\langle C\right\rangle $
versus $q$ for SPL which proves that it is logical to apply
similarity network to evolutionary study. Then, we further
investigate the distribution of sequences by measuring the
topology of similarity network. And the results are listed as
follows. Far below the threshold, the values of both $var(d_{ij})$
and $\left\langle C\right\rangle $ are independent of $q$. The
distribution is uniform in the sequence space. Near the
threshold,\ the sequences have a large variance of Hamming
distance $var(d_{ij})$ and a large clustering coefficient $C$ of
the network, which implies that the scattered band of the
distribution is wide and the cliques have formed in the
distribution of sequences, respectively. So, we have concluded
that there are several local optima around the center (global
optimum) in the sequence space. Far above the threshold, the
sequences have a small variance of Hamming distance $var(d_{ij})$
but a large clustering coefficient $C$. So the range of the
distribution will reduce to one point in the sequence space if q
approaches to 1. Finally, it has been found that $C(k)$ follows
lognormal distribution and the curve of $C$ versus $d_{0}$ is seen
as a straight line near the threshold.

Our study first applies similarity network to study the Eigen model. So, we
hope that our work can be served as motivation and a source of inspiration for
uncovering the mechanism of evolution. Moreover, finding a suitable way to
offer a bench mark for evaluating various fitness landscapes to estimate how
the real-valued landscapes is on earth will be considered in forthcoming work.

\section*{Acknowledgments} This paper was supported by the National
Science Foundation of China under Grant No. 10105007.

\newpage
\begin{figure}[tbp]
\begin{center}
{\includegraphics[ natheight=3.200700in, natwidth=4.065500in,
height=3.3892in, width=4.2973in
]%
{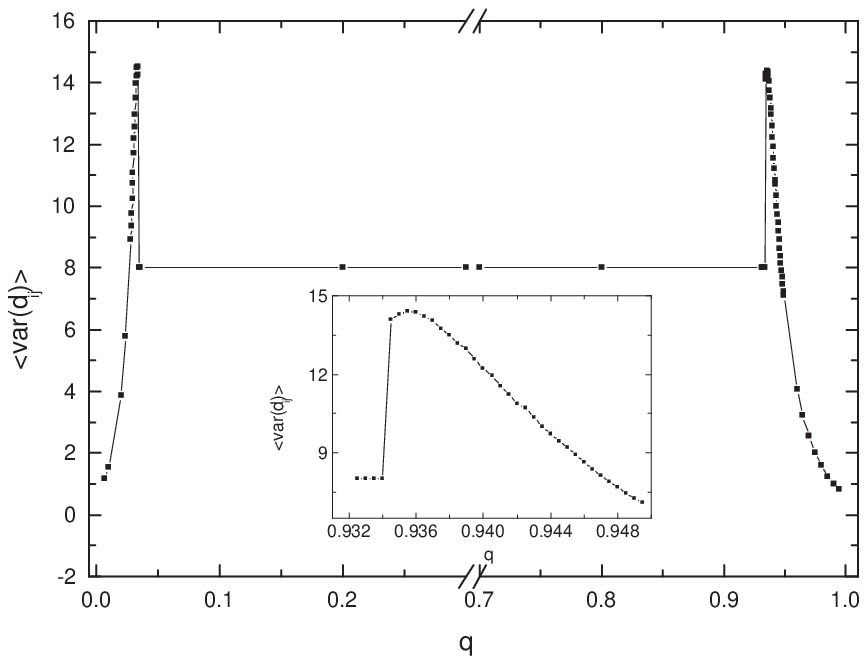}%
}%
\end{center}
\caption{The average of the variance of Hamming distance of
sequences $\left\langle var(d_{ij})\right\rangle $ versus copying
fidelity $q$ in SPL. $\left\langle ...\right\rangle $ represents
the average over 1000 evolution steps from $t=9000$ to $t=10000$
for the given $q.$ The peaks of the curve are at $q=0.9355$ and
$q=0.0336$. }
\end{figure}

\begin{figure}[tbp]
\begin{center}%
{\includegraphics[ natheight=3.200700in, natwidth=4.080200in,
height=3.3892in, width=4.312in
]%
{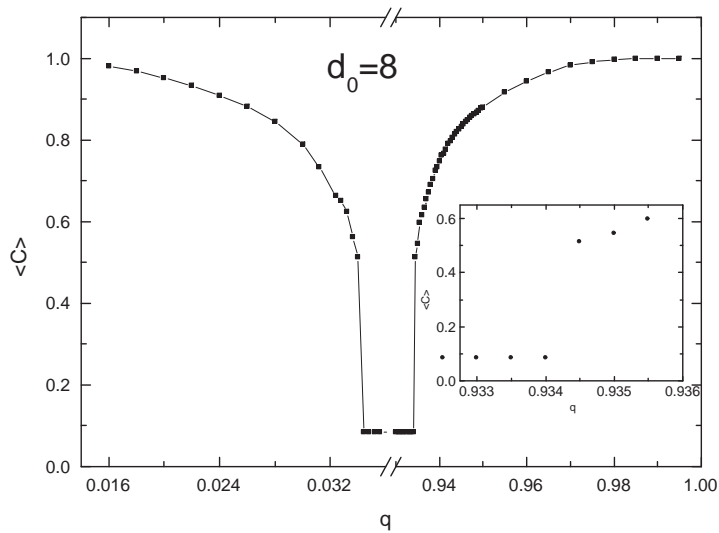}%
}%

\caption{The average of clustering coefficient $\left\langle
C\right\rangle$ of the similarity network with $d_{0}=8$
$\left\langle C\right\rangle $ versus copying fidelity $q$ in SPL.
$\left\langle ...\right\rangle $ represents the average over 1000
evolution steps from $t=9000$ to $t=10000$ of the similarity
network with $d_{0}=8$ for a given $q.$ The jump discontinuities
are at $q=0.934$ and $q=0.0344$, the same as those in figure 1.}
\end{center}
\end{figure}

\begin{figure}[tbp]
\begin{center}%
{\includegraphics[ natheight=3.2in, natwidth=4.4in,
height=2.4284in, width=3.307in
]%
{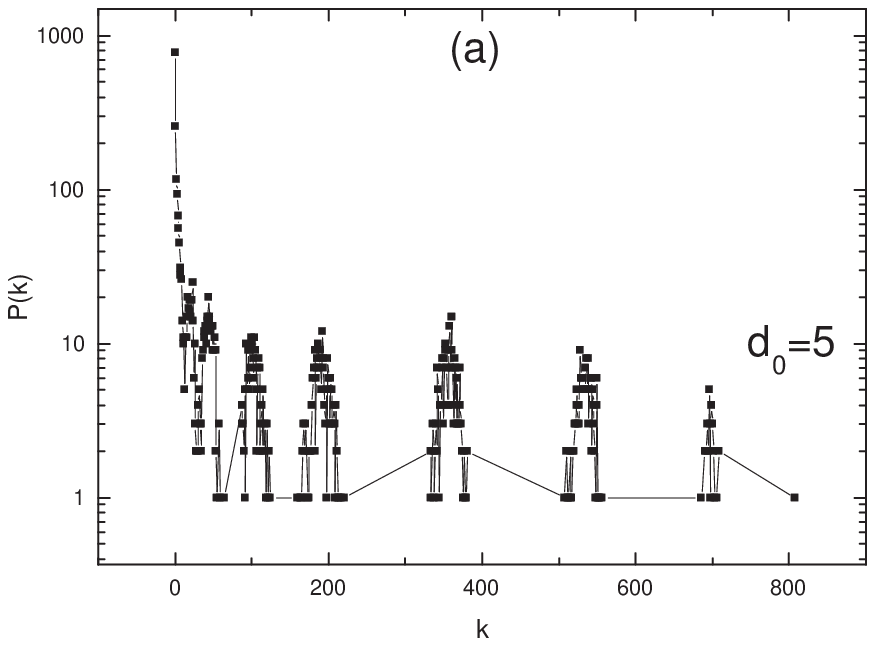}%
}%
{\includegraphics[ natheight=3.2in, natwidth=4.in,
height=2.4284in, width=3.0761in
]%
{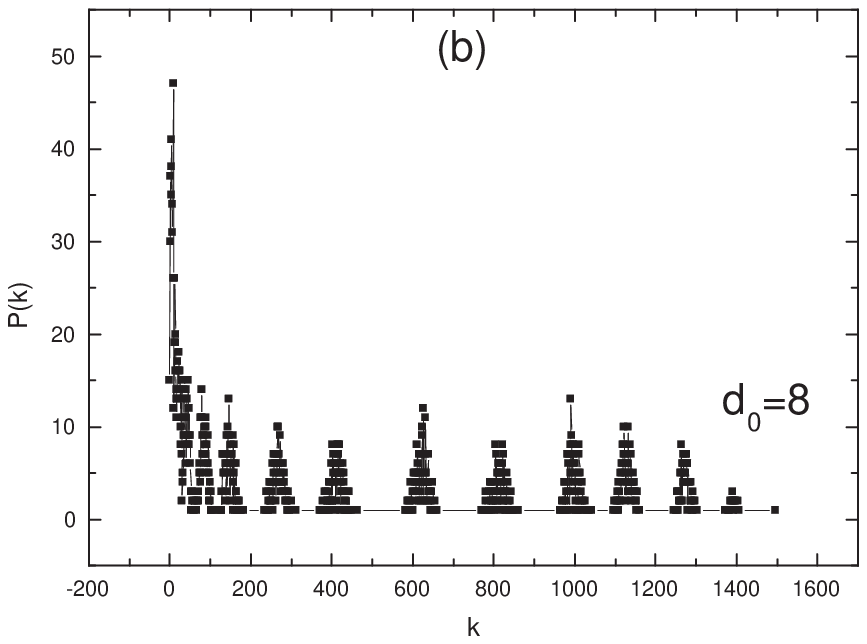}%
}%
\\
{\includegraphics[ natheight=3.200700in, natwidth=4.080200in,
height=2.4284in, width=3.0874in
]%
{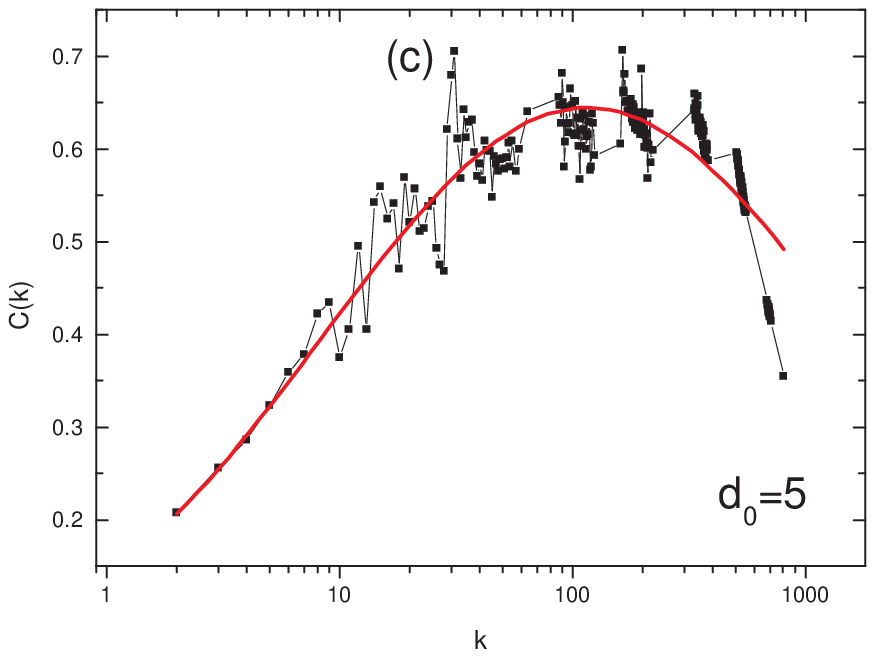}%
}%
{\includegraphics[ natheight=3.200700in, natwidth=4.080200in,
height=2.4284in, width=3.0874in
]%
{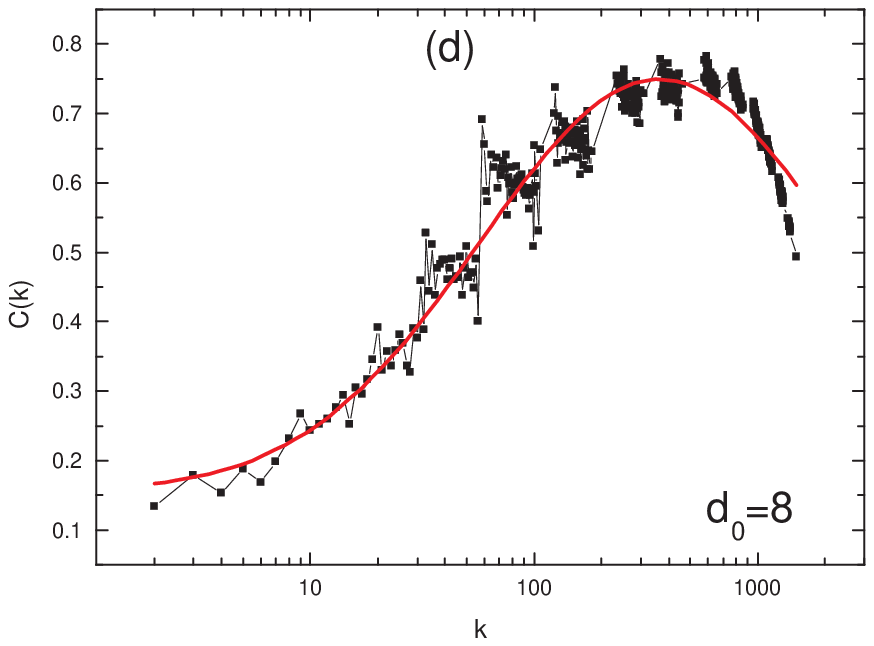}%
}%
\caption{(a), (b): The degree distribution of the similarity
network constructed in SPL with $q=0.9355$ at $t=10000$. (c), (d):
The clustering coefficient $C(k)$ versus $k$ of the same
similarity network. The data of degree $k$ have been binned
logarithmically. The curves of $C(k)$ versus $k$ follow lognormal
distribution.}

\end{center}
\end{figure}

\begin{figure}[tbp]
\begin{center}%
{\includegraphics[ natheight=3.200700in, natwidth=4.122600in,
height=2.4284in, width=3.1194in
]%
{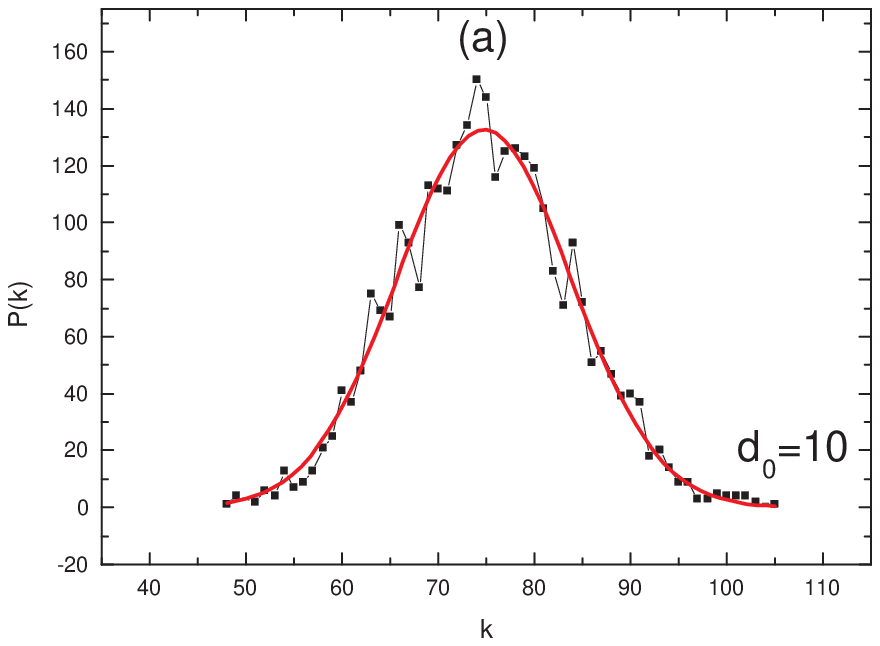}%
}%
{\includegraphics[ natheight=3.200700in, natwidth=4.135500in,
height=2.4284in, width=3.1298in
]%
{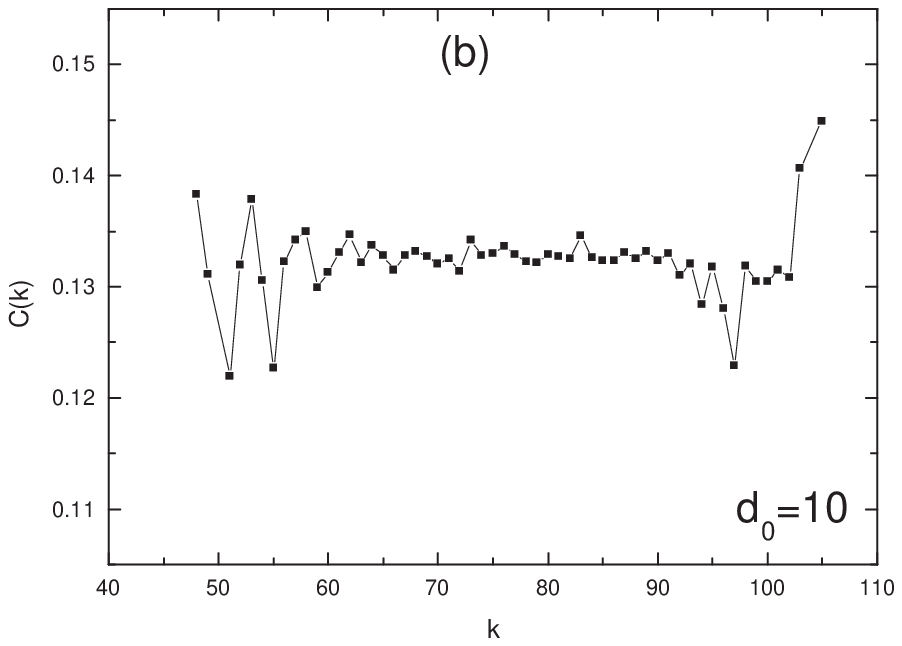}%
}%
\end{center}
\caption{(a): The degree distribution of the similarity network
with $d_{0}=10$ constructed in SPL with $q=0.6$ at $t=10000.$ (b):
The clustering coefficient $C(k)$ versus degree $k$ of the same
similarity network. The degree distribution follows Gauss
distribution when $d_{0}$ is large than 7. All the curves of
$C(k)$ approximate to a horizontal line. And when $d_{0}$ is less
than or equal to 4, the value of $C(k)$ is equal to 0.}

\end{figure}

\begin{figure}[hptb]
\begin{center}%
{\includegraphics[ natheight=3.200700in, natwidth=4.122600in,
height=2.4284in, width=3.1194in
]%
{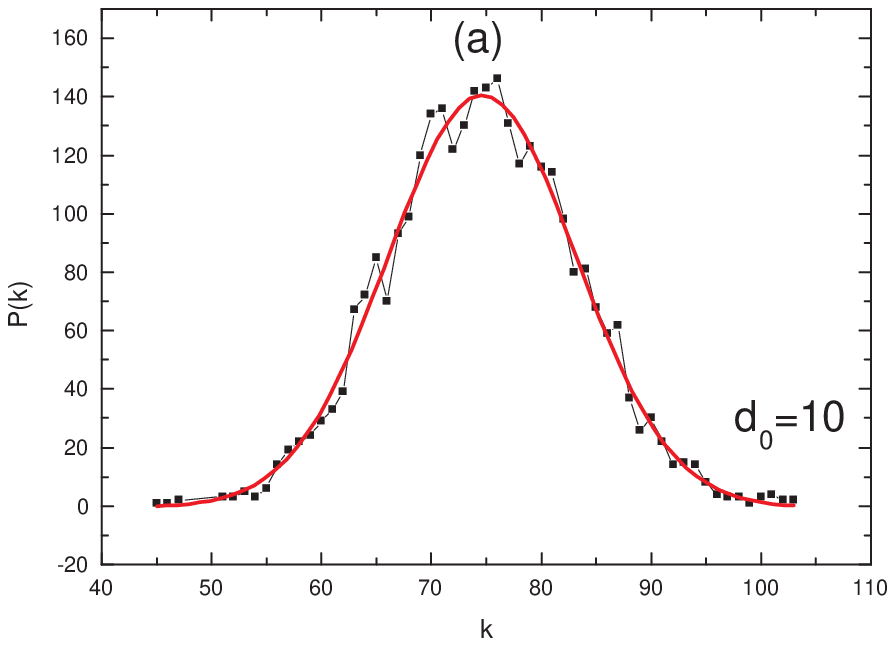}%
}%
{\includegraphics[ natheight=3.200700in, natwidth=4.135500in,
height=2.4284in, width=3.1298in
]%
{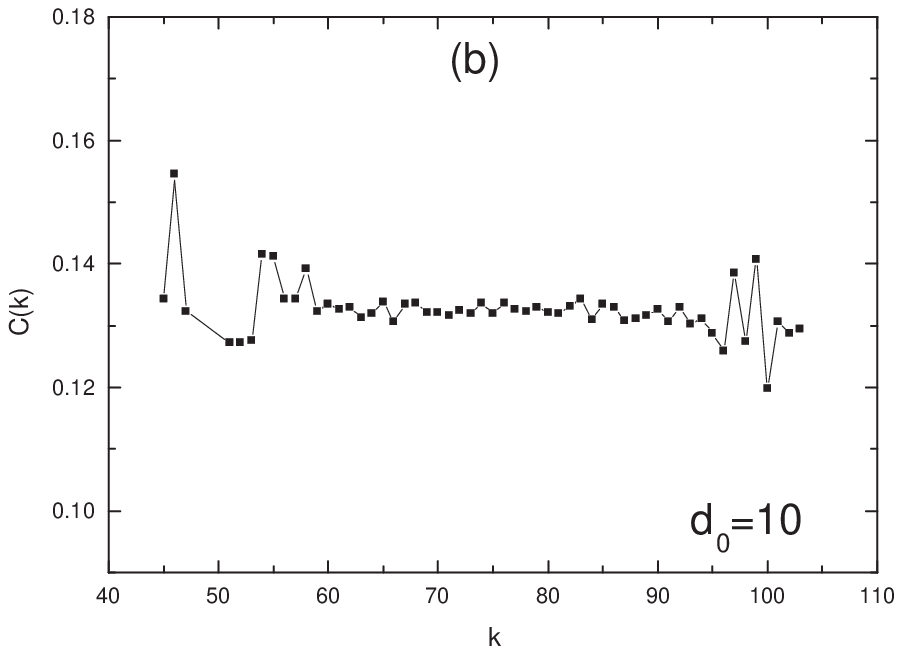}%
}%
\end{center}
\caption{(a): The degree distribution of the similarity network
with $d_{0}=10$ constructed by random sequences. (b): The
clustering coefficient $C(k)$ versus degree $k$ of the same
similarity network. The degree distribution follows Gauss
distribution when $d_{0}$ is large than 7. All the curves of
$C(k)$ approximate to a horizontal line. And when $d_{0}$ is less
than or equal to 4, the value of $C(k)$ is equal to 0.}
\end{figure}

\begin{figure}[ptb]
{\includegraphics[ natheight=3.325200in, natwidth=4.080200in,
height=2.5218in, width=3.0874in
]%
{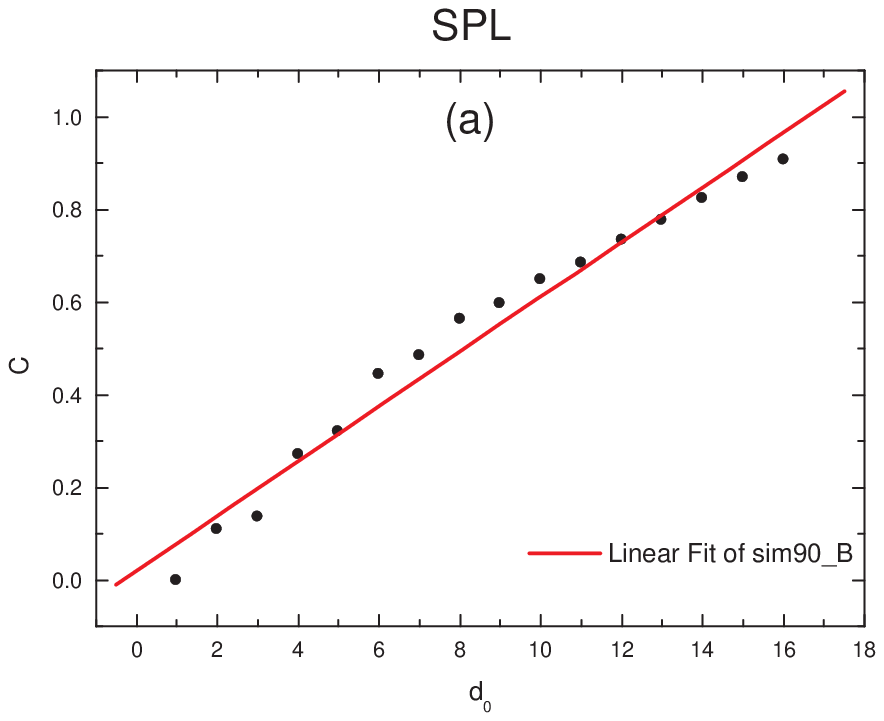}%
}%
{\includegraphics[ natheight=3.298400in, natwidth=4.177900in,
height=2.5019in, width=3.1618in
]%
{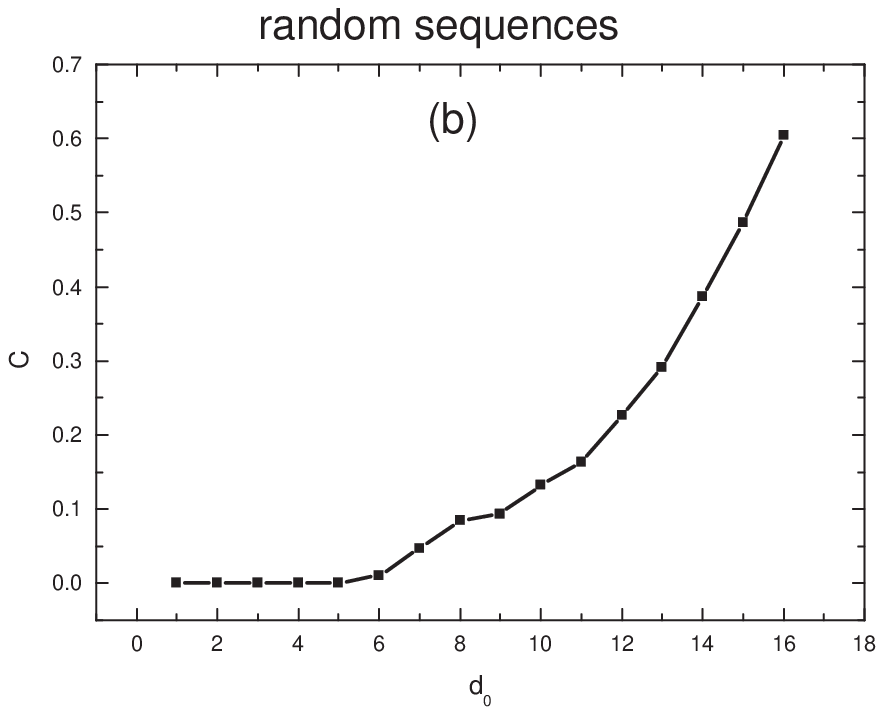}%
}%

\caption{(a): Clustering coefficient $C$ of the similarity network
for $q=0.9355$ at $t=10000$ in SPL versus Hamming distance
$d_{0}.$ (b): Clustering coefficient $C$ of the similarity network
for random sequences versus Hamming distance $d_{0}$}

\end{figure}

\end{document}